\documentclass[aps,prl,twocolumn,showpacs,superscriptaddress,groupedaddress]{revtex4}

\usepackage{graphicx} 
\usepackage{dcolumn}  
\usepackage{bm}       
\usepackage{multirow} 
\usepackage{subfigure}
\usepackage{amsmath}
\usepackage{amssymb}   
\usepackage{lineno}
\usepackage{array}
\usepackage{xcolor}

\begin{document}


\begin{widetext}

\hspace{5.2in} \mbox{Fermilab-Pub-14/309-E}

\title{Measurement of the direct CP-violating parameter $\bm{A_{\text{CP}}}$ in the decay $\bm{D^+ \to K^-\pi^+\pi^+}$}
\affiliation{LAFEX, Centro Brasileiro de Pesquisas F\'{i}sicas, Rio de Janeiro, Brazil}
\affiliation{Universidade do Estado do Rio de Janeiro, Rio de Janeiro, Brazil}
\affiliation{Universidade Federal do ABC, Santo Andr\'e, Brazil}
\affiliation{University of Science and Technology of China, Hefei, People's Republic of China}
\affiliation{Universidad de los Andes, Bogot\'a, Colombia}
\affiliation{Charles University, Faculty of Mathematics and Physics, Center for Particle Physics, Prague, Czech Republic}
\affiliation{Czech Technical University in Prague, Prague, Czech Republic}
\affiliation{Institute of Physics, Academy of Sciences of the Czech Republic, Prague, Czech Republic}
\affiliation{Universidad San Francisco de Quito, Quito, Ecuador}
\affiliation{LPC, Universit\'e Blaise Pascal, CNRS/IN2P3, Clermont, France}
\affiliation{LPSC, Universit\'e Joseph Fourier Grenoble 1, CNRS/IN2P3, Institut National Polytechnique de Grenoble, Grenoble, France}
\affiliation{CPPM, Aix-Marseille Universit\'e, CNRS/IN2P3, Marseille, France}
\affiliation{LAL, Universit\'e Paris-Sud, CNRS/IN2P3, Orsay, France}
\affiliation{LPNHE, Universit\'es Paris VI and VII, CNRS/IN2P3, Paris, France}
\affiliation{CEA, Irfu, SPP, Saclay, France}
\affiliation{IPHC, Universit\'e de Strasbourg, CNRS/IN2P3, Strasbourg, France}
\affiliation{IPNL, Universit\'e Lyon 1, CNRS/IN2P3, Villeurbanne, France and Universit\'e de Lyon, Lyon, France}
\affiliation{III. Physikalisches Institut A, RWTH Aachen University, Aachen, Germany}
\affiliation{Physikalisches Institut, Universit\"at Freiburg, Freiburg, Germany}
\affiliation{II. Physikalisches Institut, Georg-August-Universit\"at G\"ottingen, G\"ottingen, Germany}
\affiliation{Institut f\"ur Physik, Universit\"at Mainz, Mainz, Germany}
\affiliation{Ludwig-Maximilians-Universit\"at M\"unchen, M\"unchen, Germany}
\affiliation{Panjab University, Chandigarh, India}
\affiliation{Delhi University, Delhi, India}
\affiliation{Tata Institute of Fundamental Research, Mumbai, India}
\affiliation{University College Dublin, Dublin, Ireland}
\affiliation{Korea Detector Laboratory, Korea University, Seoul, Korea}
\affiliation{CINVESTAV, Mexico City, Mexico}
\affiliation{Nikhef, Science Park, Amsterdam, the Netherlands}
\affiliation{Radboud University Nijmegen, Nijmegen, the Netherlands}
\affiliation{Joint Institute for Nuclear Research, Dubna, Russia}
\affiliation{Institute for Theoretical and Experimental Physics, Moscow, Russia}
\affiliation{Moscow State University, Moscow, Russia}
\affiliation{Institute for High Energy Physics, Protvino, Russia}
\affiliation{Petersburg Nuclear Physics Institute, St. Petersburg, Russia}
\affiliation{Instituci\'{o} Catalana de Recerca i Estudis Avan\c{c}ats (ICREA) and Institut de F\'{i}sica d'Altes Energies (IFAE), Barcelona, Spain}
\affiliation{Uppsala University, Uppsala, Sweden}
\affiliation{Taras Shevchenko National University of Kyiv, Kiev, Ukraine}
\affiliation{Lancaster University, Lancaster LA1 4YB, United Kingdom}
\affiliation{Imperial College London, London SW7 2AZ, United Kingdom}
\affiliation{The University of Manchester, Manchester M13 9PL, United Kingdom}
\affiliation{University of Arizona, Tucson, Arizona 85721, USA}
\affiliation{University of California Riverside, Riverside, California 92521, USA}
\affiliation{Florida State University, Tallahassee, Florida 32306, USA}
\affiliation{Fermi National Accelerator Laboratory, Batavia, Illinois 60510, USA}
\affiliation{University of Illinois at Chicago, Chicago, Illinois 60607, USA}
\affiliation{Northern Illinois University, DeKalb, Illinois 60115, USA}
\affiliation{Northwestern University, Evanston, Illinois 60208, USA}
\affiliation{Indiana University, Bloomington, Indiana 47405, USA}
\affiliation{Purdue University Calumet, Hammond, Indiana 46323, USA}
\affiliation{University of Notre Dame, Notre Dame, Indiana 46556, USA}
\affiliation{Iowa State University, Ames, Iowa 50011, USA}
\affiliation{University of Kansas, Lawrence, Kansas 66045, USA}
\affiliation{Louisiana Tech University, Ruston, Louisiana 71272, USA}
\affiliation{Northeastern University, Boston, Massachusetts 02115, USA}
\affiliation{University of Michigan, Ann Arbor, Michigan 48109, USA}
\affiliation{Michigan State University, East Lansing, Michigan 48824, USA}
\affiliation{University of Mississippi, University, Mississippi 38677, USA}
\affiliation{University of Nebraska, Lincoln, Nebraska 68588, USA}
\affiliation{Rutgers University, Piscataway, New Jersey 08855, USA}
\affiliation{Princeton University, Princeton, New Jersey 08544, USA}
\affiliation{State University of New York, Buffalo, New York 14260, USA}
\affiliation{University of Rochester, Rochester, New York 14627, USA}
\affiliation{State University of New York, Stony Brook, New York 11794, USA}
\affiliation{Brookhaven National Laboratory, Upton, New York 11973, USA}
\affiliation{Langston University, Langston, Oklahoma 73050, USA}
\affiliation{University of Oklahoma, Norman, Oklahoma 73019, USA}
\affiliation{Oklahoma State University, Stillwater, Oklahoma 74078, USA}
\affiliation{Brown University, Providence, Rhode Island 02912, USA}
\affiliation{University of Texas, Arlington, Texas 76019, USA}
\affiliation{Southern Methodist University, Dallas, Texas 75275, USA}
\affiliation{Rice University, Houston, Texas 77005, USA}
\affiliation{University of Virginia, Charlottesville, Virginia 22904, USA}
\affiliation{University of Washington, Seattle, Washington 98195, USA}
\author{V.M.~Abazov} \affiliation{Joint Institute for Nuclear Research, Dubna, Russia}
\author{B.~Abbott} \affiliation{University of Oklahoma, Norman, Oklahoma 73019, USA}
\author{B.S.~Acharya} \affiliation{Tata Institute of Fundamental Research, Mumbai, India}
\author{M.~Adams} \affiliation{University of Illinois at Chicago, Chicago, Illinois 60607, USA}
\author{T.~Adams} \affiliation{Florida State University, Tallahassee, Florida 32306, USA}
\author{J.P.~Agnew} \affiliation{The University of Manchester, Manchester M13 9PL, United Kingdom}
\author{G.D.~Alexeev} \affiliation{Joint Institute for Nuclear Research, Dubna, Russia}
\author{G.~Alkhazov} \affiliation{Petersburg Nuclear Physics Institute, St. Petersburg, Russia}
\author{A.~Alton$^{a}$} \affiliation{University of Michigan, Ann Arbor, Michigan 48109, USA}
\author{A.~Askew} \affiliation{Florida State University, Tallahassee, Florida 32306, USA}
% ...
%
%
\collaboration{The D0 Collaboration\footnote{with visitors from
$^{a}$Augustana College, Sioux Falls, SD, USA,
$^{b}$The University of Liverpool, Liverpool, UK,
$^{c}$DESY, Hamburg, Germany,
$^{d}$Universidad Michoacana de San Nicolas de Hidalgo, Morelia, Mexico
$^{e}$SLAC, Menlo Park, CA, USA,
$^{f}$University College London, London, UK,
$^{g}$Centro de Investigacion en Computacion - IPN, Mexico City, Mexico,
$^{h}$Universidade Estadual Paulista, S\~ao Paulo, Brazil,
$^{i}$Karlsruher Institut f\"ur Technologie (KIT) - Steinbuch Centre for Computing (SCC),
D-76128 Karlsruhe, Germany,
$^{j}$Office of Science, U.S. Department of Energy, Washington, D.C. 20585, USA,
$^{k}$American Association for the Advancement of Science, Washington, D.C. 20005, USA,
$^{l}$Kiev Institute for Nuclear Research, Kiev, Ukraine,
$^{m}$University of Maryland, College Park, Maryland 20742, USA
and
$^{n}$European Orgnaization for Nuclear Research (CERN), Geneva, Switzerland
}} \noaffiliation
\vskip 0.25cm

\date{August 28, 2014}

\begin{abstract}
We measure the direct CP-violating parameter $A_{\text{CP}}$ for the decay of the charged charm meson, $D^+ \to K^-\pi^+\pi^+$ (and charge conjugate), using the full 10.4~fb$^{-1}$ sample of $p\bar{p}$ collisions at $\sqrt{s}=1.96$~TeV collected by the D0 detector at the Fermilab Tevatron collider. We extract the raw reconstructed charge asymmetry by fitting the invariant mass distributions for the sum and difference of charge-specific samples. This quantity is then corrected for detector-related asymmetries using data-driven methods and for possible physics asymmetries (from $B \to D$ processes) using input from Monte Carlo simulation. We measure $A_{\text{CP}} = [-0.16 \pm 0.15 \text{\,(stat.)} \pm 0.09 \text{\,(syst.)}]\%$, which is consistent with zero, as expected from the standard model prediction of CP conservation, and is the most precise measurement of this quantity to date.
\end{abstract}

\maketitle

\end{widetext}

The violation of CP symmetry in the fundamental interactions of particle physics is required to explain the matter dominance of the universe~\cite{sakharov,theo1,theo2}. The standard model (SM) describes CP violation in the quark sector through the presence of a single complex phase in the CKM matrix. This matrix dictates the strength of flavor transitions through the weak interaction. All experimental observations to date are consistent with a single complex phase~\cite{pdg}, with the exception of a small number of discrepancies at the $\approx$3$\sigma$ level, most notably the anomalously large same-charge dimuon asymmetry measurement from the D0 experiment~\cite{dimuon2014}. However, the degree of CP violation in the SM is insufficient to explain the cosmological matter dominance~\cite{huet}. 
It is therefore important to continue searching for sources of CP violation beyond those predicted by the SM.

Decays of heavy-flavor hadrons provide a natural testing ground for these searches. In particular, decays proceeding through box or penguin diagrams are highly sensitive to possible CP violation contributions from processes beyond the SM induced by additional particles in the loops. 
However, due to the difficulty in simultaneously extracting production, detection and physics asymmetries, these searches for anomalous CP violation typically measure the difference in charge asymmetries between the channel of interest and a Cabibbo-favored reference channel, which is then assumed to be CP symmetric~\cite{acpbplus_lhcb, assl, assl_lhcb, adsl, lhcb_deltaacp}.
Performing high-precision measurements of CP violation parameters in these Cabibbo-favored decays is therefore crucial in order to establish an experimental basis for these assumptions, thus reducing dependence on theoretical predictions. 
The data set collected by the D0 experiment at the Tevatron $p\bar{p}$ collider is uniquely suited to perform such measurements, having a CP-symmetric initial state, a charge-symmetric tracking detector, and almost equal beam exposure in all four combinations of solenoid and toroid magnet polarities.

In this Letter, we describe the measurement of the direct CP violation parameter in the Cabibbo-favored decay $D^+ \to K^-\pi^+\pi^+$ (charge conjugate states are implied throughout this paper), defined as
\begin{eqnarray}
&&A_{\text{\text{CP}}}(D^+ \to K^-\pi^+\pi^+)  =  \\
&&\frac{\Gamma(D^{+} \to K^-\pi^+\pi^+) - \Gamma(D^{-} \to K^+\pi^-\pi^-)} 
                                               {\Gamma(D^{+} \to K^-\pi^+\pi^+) + \Gamma(D^{-} \to K^+\pi^-\pi^-)},\nonumber
\end{eqnarray}
and hereafter denoted $A_{\text{\text{CP}}}$.
Currently this parameter has only been measured by the CLEO collaboration~\cite{cleo_2014}: 
$A_{\text{\text{CP}}} = [ -0.3 \pm 0.2 \text{\,(stat.)} \pm 0.4 \text{\,(syst.)}]\%$.
We use the complete sample of $p\bar{p}$ collisions generated by the Tevatron accelerator at $\sqrt{s} = 1.96$~TeV and collected by the D0 detector. This corresponds to approximately $10.4$~fb$^{-1}$ of integrated luminosity.

CP violation can only occur if there is interference between two amplitudes with different strong and weak phases. For the decay mode being investigated, this requirement is not satisfied, with two tree-level amplitudes both proportional to the product of CKM matrix elements $V_{cs}^*V_{ud}$ and no contribution from Cabibbo-supressed diagrams. The SM therefore predicts negligible CP violation with respect to the experimental uncertainties. Any significant deviation of $A_{\text{\text{CP}}}$ from zero would thus constitute evidence for new physics contributions~\cite{silva}.

Experimentally, the CP asymmetry parameter is determined by measuring a raw charge asymmetry ($A$) and applying corrections to account for differences in the detection of the final state particles ($A_{\text{det}}$) and in the production rates of $D^+$ and $D^-$ mesons ($A_{\text{phys}}$), i.e.,
\begin{eqnarray}
A_{\text{\text{CP}}}  &=& A - A_{\text{det}} - A_{\text{phys}}. \label{eq:main}
\end{eqnarray}
The raw quantity $A$ is the asymmetry in the number of $D^+$ versus $D^-$ mesons reconstructed in the described decay mode and passing all selection requirements. It is extracted by simultaneously fitting the $M(K\pi\pi)$ invariant mass distributions for the sum of all candidates and for the difference $N(D^+) - N(D^-)$. The detector asymmetry $A_{\text{det}}$ accounts for differences in the reconstruction efficiency for positive and negative kaons, pions, and muons and is determined using methods based on data in dedicated independent channels. The physics asymmetry $A_{\text{phys}}$ accounts for possible charge-asymmetric production of $D$ mesons arising through the decay of $B$ hadrons. For each possible source, the contribution to $A_{\text{phys}}$ is the product of the relevant CP asymmetry (taken from the world-average of experimental results) and the fraction of $D$ mesons arising from this source (determined from simulation).
In practice the value of $A_{\text{phys}}$ is small compared to the precision of the final measurement, while the detector correction is significant.
For simplicity, we use $D$ to collectively denote $D^{\pm}$ mesons throughout this paper. In cases where distinguishing the charge is important we explicitly include it.

The D0 detector is described in detail elsewhere~\cite{d0det1,d0det2}. The most important components for this analysis are the central tracking detector, the muon system, and the magnets. The central tracking system comprises a silicon microstrip tracker (SMT) closest to the beampipe, surrounded by a central fiber tracker (CFT), with the entire system located within a 1.9~T solenoidal field. The SMT (CFT) has polar acceptance $|\eta| < 3$ ($|\eta| < 2.5$), where the pseudorapidity is defined as $\eta = -\text{ln}[\text{tan}(\theta/2)$], and $\theta$ is the polar angle with respect to the positive $z$ axis along the proton beam direction. The muon system (covering $|\eta|<2$) comprises a layer of tracking detectors and scintillation trigger counters in front of 1.8~T toroid magnets, followed by two similar layers after the toroids. The polarities of both the solenoid and toroid magnets were regularly reversed approximately every two weeks during data collection to give near equal exposure in all four configurations.
The magnet reversal ensures that the main detector asymmetries cancel to first order by symmetrizing the detector acceptance for positive and negative particles. The residual deviations from equal exposure (typically less than 5\%) are removed by weighting events according to their polarity to force equal contributions from all four polarity configurations.

In the absence of a dedicated trigger for hadronic decays of heavy-flavor hadrons, we use a suite of single muon and dimuon triggers to select the data sample, along with an offline single muon filter. Events that exclusively satisfy triggers using track impact parameter information are removed to avoid lifetime biases which influence the $D$ meson parentage, and which are challenging to model in simulation. The muon trigger and offline requirements can bias the composition of the data in favor of semileptonic decays of charm and bottom hadrons. In particular, the fraction of $D$ mesons arising from semileptonic decays of $B$ mesons will be enhanced. These requirements must be taken into account when determining both detector and physics asymmetry corrections. To facilitate this process, the analysis places particular requirements on the muon quality and kinematic variables, to match those used when determining kaon, pion, and muon reconstruction asymmetries. The muon must produce hits in the muon tracking layers both inside and outside the toroid, and must be spatially-matched to a central track with total momentum $p(\mu) > 3$~GeV/$c$ and transverse momentum $p_T(\mu) > 2$~GeV/$c$. The selected muon is not used in the subsequent reconstruction of $D$ meson signal candidates. In particular, no further requirements are imposed which use the muon information (for example, charge, or spatial origin with respect to the $D$ meson candidate).

For events passing the muon selection, $D$ candidates are reconstructed from all possible three-track combinations that have total charge $q=\pm 1$ and that are consistent with arising from a common vertex. The three tracks must satisfy quality requirements and each track must have $p_T>0.7$~GeV/$c$. The two like-charge tracks are assigned the charged pion mass, and the third track is assigned the charged kaon mass~\cite{pdg}. The resulting invariant mass of the $D$ candidate must lie within $1.65 < M(K\pi\pi) < 2.05$~GeV/$c^2$, and the momentum and displacement vectors of the reconstructed $D$ meson must point in the same hemisphere. Additionally, the transverse decay length of the $D$ candidate must exceed three times its uncertainty, $L_{xy}(D)/\sigma[L_{xy}(D)] > 3$. The transverse decay length is defined as the displacement between the $p\bar{p}$ primary interaction vertex and the reconstructed $D$ meson decay vertex, projected onto the plane perpendicular to the beam direction.

The final selection of events uses a log-likelihood ratio (LLR) method to combine twelve individual variables into a single multivariate discriminant, using a similar approach to that described in Ref.~\cite{adsl}. 
The input variables are as follows: the transverse momenta of the three final-state hadrons and their track isolations, the transverse decay length of the $D$ meson $L_{xy}$ and its significance $L_{xy}(D)/\sigma[L_{xy}(D)]$, the $\chi^2$ of the vertex fit of the three tracks, the angular separations of the kaon and lowest-$p_T$ pion and of the two pions, and the cosine of the angle between the momentum and displacement vectors of the $D$ meson candidate. The angular separation of two tracks is defined as $\Delta R = \sqrt{\Delta\phi^2 + \Delta\eta^2}$, where $\Delta\phi$ and $\Delta\eta$ are the track separations in the azimuthal angle and pseudorapidity, respectively.
The track isolation $I$ is the momentum of a particle divided by the sum of the momenta of all tracks contained in a cone of size $\Delta R<0.5$ around the particle. Tracks corresponding to the other two final state particles for this candidate are excluded from the sum.

The background distributions used to construct the LLR discriminant are populated using 1\% of the data, chosen by randomly sampling the $D$ candidates following all requirements except for the LLR. This sample has a small signal contamination (around 0.4\%), but it is found to provide the best overall discriminant performance. No correction is applied to account for the negligible effect of real signal events in this sample. The signal distributions are modeled using Monte Carlo (MC) simulation of inclusive $D^{\pm} \to K^{\mp}\pi^{\pm}\pi^{\pm}$ events, without any constraints on their origin. The final requirement on the LLR output is chosen to maximize the signal significance in the 1\% random data sample (scaling-up to extrapolate to the full sample). Ensemble studies confirm that this corresponds to the minimum uncertainty on the final asymmetry measurement.

For all simulated samples, events are generated using {\sc pythia} version 6.409~\cite{pythia} interfaced with {\sc evtgen}~\cite{evtgen} to model the decays of particles containing $b$ and $c$ quarks. The generation model includes all quark flavors, ensuring that charm and bottom quarks from gluon splitting are properly included in the final sample. Generated events are processed by a {\sc geant}-based detector simulation~\cite{geant}, overlaid with data from randomly collected bunch crossings to simulate pile-up from multiple interactions, and reconstructed using the same software as used for data.

\begin{figure}[t]
        \centering
        \includegraphics[width=\columnwidth]{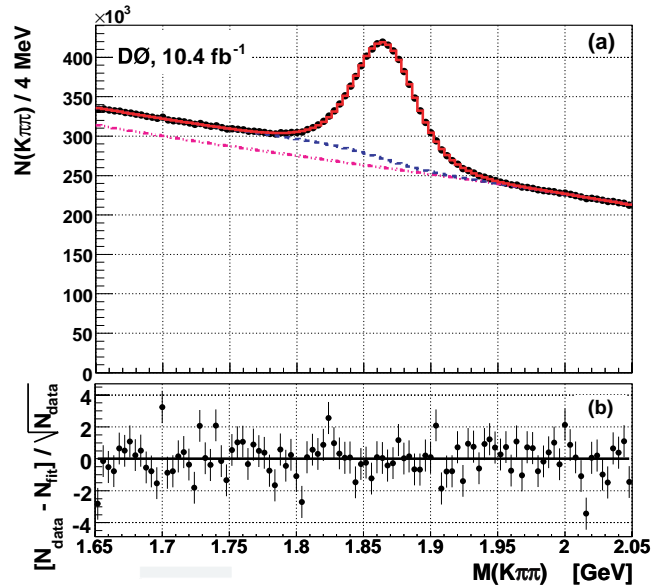}
        \caption[]{(a) Invariant mass distribution $M(K\pi\pi)$ after all selections have been applied (data markers). Also shown is the result of the fit to the data, as described in the text (solid line). To illustrate the contributions of the three separate components, the total background (dashed line) and polynomial function (dot-dashed line) are shown separately. (b) Fit residuals $[N_{\text{data}} - N_{\text{fit}}] / \sqrt{N_{\text{data}}}$, demonstrating the agreement between the data and the fit model.}
\label{fig:fit_sum}
\end{figure}

The $M(K\pi\pi)$ distribution of candidates passing all selection requirements is shown in Fig.~\ref{fig:fit_sum}, along with the results of a fit to the data (described later). A total of approximately 31 million candidates is found, of which $N(D^{\pm}) = 2\,270\,224 \pm 7\,406$ are assigned as $D^{\pm}$ signal in the fit. The effective statistical loss caused by the magnet polarity weighting, included in this number, is 3.2\%.

\begin{figure}[t]
        \centering
        \includegraphics[width=\columnwidth]{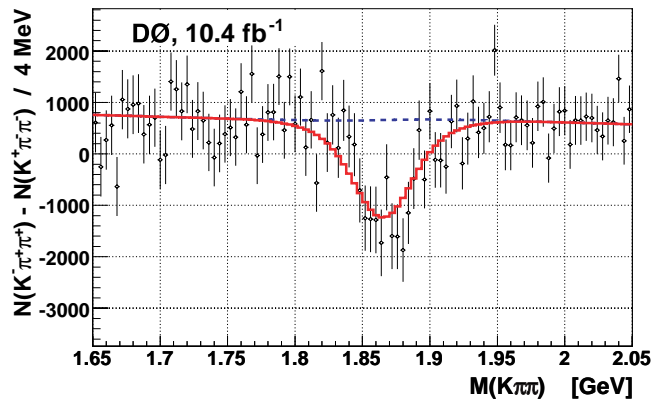}
        \caption[]{Invariant mass distribution $M(K\pi\pi)$ for the difference $N(D^+) - N(D^-)$ (data markers). Also shown are the result of the fit to the data (solid line) and the overall background contribution (dashed line). 
}
\label{fig:fit_diff}
\end{figure}

The raw asymmetry $A$ is extracted through a simultaneous binned minimum-$\chi^2$ fit of the sum distribution (in Fig.~\ref{fig:fit_sum}) and the difference distribution $[N(D^+) - N(D^-)]$ (in Fig.~\ref{fig:fit_diff}). The method is the same as described in Ref.~\cite{adsl}, with the only difference being a slight simplification of the combinatorial background model, enabled by the updated event selection criteria. The fit includes three components, each set to have the same shape in the sum and difference distributions, with only their relative normalizations differing in the two cases. The $D$ signal is parametrized by two Gaussian functions constrained to have the same mean value, to model the effect of the detector mass resolution. A hyperbolic tangent function is used to model the effect of a range of multi-body physics backgrounds, including both partially reconstructed decays of $D^{(*)}$ mesons, and reflections where the final-state hadrons are assigned the wrong mass. 
The main contributions are from $D^{+}$ decays to $K^{-}\pi^{+}\pi^{+}\pi^0$, $\pi^{-}\pi^{+}\pi^{+}\pi^0$, and $K^{-}K^{+}\pi^{+}$; $D_s^+$ decays to $K^+K^-\pi^+$; $\bar{D}^0$ decays to four charged hadrons; and decays of $D^{*+} \to D^0\pi^{+}$, with $D^0 \to K^{-}\pi^{+}\pi^0$, where in all cases the $\pi^0$ is not reconstructed. 
The hyperbolic tangent parametrization is chosen based on studies of decay-specific and inclusive simulated samples and is the same as used in Ref.~\cite{adsl}. The inflection-point is fixed for the nominal fit, based on simulation, but is allowed to  vary when assigning a systematic uncertainty to the choice of fitting model. The steepness of the slope is constrained based on the resolution of the Gaussian peak in data~\cite{adsl}, which is also well-motivated by simulation. Finally, the smooth combinatorial background is modeled by a polynomial with constant, linear, and cubic terms. The quadratic term is excluded since it does not improve the goodness-of-fit. For the fit to the difference distribution, the relative contributions of the three components are quantified through asymmetry parameters, including the raw asymmetry $A$ for the signal and corresponding asymmetries $A_{\text{multi}}$ and $A_{\text{comb}}$ for the multi-body and combinatorial components, respectively. Hence the models used to fit the sum ($F_{\text{sum}}$) and difference ($F_{\text{diff}}$) distributions can be summarized as:
\begin{eqnarray}
F_{\text{sum}} &=& F_{\text{sig}} + F_{\text{comb}} + F_{\text{multi}}, \\
F_{\text{diff}} &=& A \cdot F_{\text{sig}} + A_{\text{comb}} \cdot F_{\text{comb}} + A_{\text{multi}} \cdot F_{\text{multi}},  \nonumber
\end{eqnarray}
where, for instance, $F_{\text{sig}}$ is the function used to model the signal component.

The total number of candidates and the difference between the positive and negative candidate counts are used as constraints to reduce the number of free parameters by two, giving improved fit stability. The final fit has ten free parameters, six for the signal (signal yield $N(D)$, invariant mass $M(D)$, the widths of the two Gaussian functions, the fraction of signal in the wider Gaussian, and the raw asymmetry) and four for the background (fraction of background in multi-body component, first- and third-order polynomial coefficients, and $A_{\text{multi}}$). The final two variables, $A_{\text{comb}}$ and the constant term in the polynomial function, are completely defined by the set of ten free parameters and the two external constraints.

The corresponding distribution and fit for the difference $[N(D^+) - N(D^-)]$ is shown in Fig.~\ref{fig:fit_diff}. A significant negative raw asymmetry is observed, $A = (-1.28 \pm 0.15)$\%, consistent with the value expected from known detector asymmetries. 
The two background asymmetries are $A_{\text{multi}} = (-0.41 \pm 0.60)$\% and $A_{\text{comb}} = (+0.27 \pm 0.04)$\%.
The main source of charge asymmetry in both background components is the kaon reconstruction asymmetry, which is around $+1.1\%$ and is described later. The sign and magnitude of both $A_{\text{multi}}$ and $A_{\text{comb}}$ are consistent with expectations from this kaon asymmetry alone. The main processes contributing to the multi-body component, and including a single charged kaon in the final state, are from the Cabibbo-favored transition $c \to s$. This results in a negative correlation between the kaon and $D$ charge, so we expect $A_{\text{multi}}$ to be negative, with a magnitude somewhat less than 1.1\% due to dilution from processes without a single charged kaon. In contrast, the combinatorial background component models the contribution of random three-track combinations: the kaon asymmetry leads to an overall excess of positive tracks, and so $A_{\text{comb}}$ is expected to be positive, with a magnitude driven by the relative abundance of kaons in the track sample.
The full fit to both distributions has a $\chi^2$ of 209 for 190 degrees-of-freedom, with no visible structures in the fit residuals and pull plots consistent with unit-width Gaussians. 

To test the sensitivity and accuracy of the fitting procedure, the data are used to create ensembles of charge-randomized pseudoexperiments with a range of different input raw asymmetries. These confirm that the asymmetry extraction is unbiased and that the statistical uncertainty reported by the fit is consistent with the expected value ($\pm 0.15\%$). Systematic uncertainties are evaluated for a range of sources by repeating the fit under several reasonable variations and examining the change in the extracted raw asymmetry. The contribution to the systematic uncertainty on $A$ from each source is taken as the RMS of the set of fit variants with respect to the nominal measurement. The upper and lower limits of the fitting range are independently varied by up to 50~MeV/$c^2$; the bin width is varied from 2 to 10~MeV/$c^2$; an alternative method is used to determine the magnet polarity weights, based on the number of fitted signal candidates (rather than the total yield) in each configuration; the combinatorial background model is varied, either by removing the cubic term, or by adding a quadratic term; and, finally, the inflection point of the hyperbolic tangent function is allowed to vary in the fit, rather than being fixed from simulation. The dominant systematic effect comes from varying the fitting range ($\pm 0.017\%$), with bin width and fitting model contributing $\pm 0.005\%$ each, and the polarity weighting method an order of magnitude smaller. The final systematic uncertainty on $A$, given by summing the individual contributions in quadrature, is $\pm 0.018\%$, much smaller than the statistical uncertainty.

The detector asymmetry has one term for each final state particle, including the implicit muon requirement, $A_{\text{det}} = 2 a^{\pi} + \rho \cdot a^{\mu} - a^{K}$, where $a^X$ is the reconstruction asymmetry for particle species $X$. The factor of 2 accounts for the two pions in the final state, and the sign of each term reflects the charge with respect to the $D$ meson.
The muon asymmetry coefficient $\rho$ is the charge correlation between the muon and $D$ meson, necessary because no explicit charge requirements are enforced in this analysis. This is extracted from the data, through separate fits of the two cases $q(\mu) \cdot q(D) = \pm 1$, yielding $\rho = -0.435 \pm 0.004$. Each of the three asymmetries $a^X$ is extracted from dedicated independent channels, and determined in appropriate kinematic bins to allow them to be applied to the signal channel by a weighted average over all bins. These input asymmetries have already been determined and documented~\cite{adsl} and used in several previous D0 publications~\cite{assl, adsl, acp_bplus, acp_ds}.

The kaon asymmetry is at least 20 times larger than all other detector effects. It arises from the larger $K^-$ cross-section with detector material than for $K^+$, leading to a higher $K^+$ reconstruction efficiency.
This asymmetry is extracted from $K^{*0} \to K^- \pi^+$ decays, in bins of absolute kaon pseudorapidity $|\eta(K)|$ and momentum $p(K)$~\cite{adsl}. Applying these to the signal sample gives a total kaon asymmetry of $a^K = (1.06 \pm 0.04 \pm 0.05)$\%. The first uncertainty is statistical, from the finite $K^{*0}$ sample size; the second uncertainty is systematic, based on variations of the $K^{*0}$ fitting method.
The pion asymmetry is investigated using $K^0_S \to \pi^+ \pi^-$ and $K^{*+} \to K^0_S \pi^+$ decays~\cite{adsl}. No indication of any asymmetry is observed, and we assign a systematic uncertainty of $\pm 0.05\%$ to account for the limited precision of this measurement.

The muon asymmetry is extracted from $J/\psi \to \mu^+ \mu^-$ decays, in bins of absolute muon pseudorapidity $|\eta(\mu)|$ and transverse momentum $p_T(\mu)$~\cite{adsl}. After convoluting the kinematically binned muon asymmetry with the corresponding signal distributions, and multiplying by the charge correlation, the final correction is $\rho \cdot a^{\mu} = (-0.045 \pm 0.011 \pm 0.004)\%$. The systematic uncertainty includes variations to the $J/\psi$ fitting procedure and to the kinematic binning scheme.
The overall detector asymmetry is then $A_{\text{det}} = [ -1.11 \pm 0.04 \text{\,(stat.)} \pm 0.07 \text{\,(syst.)} ]\%$, where statistical and systematic uncertainties from each source have been separately added in quadrature.

After correcting for detector asymmetries, we consider the asymmetry, $A_{\text{phys}}$, arising from different rates of $D^+$ and $D^-$ production.
We assume that the direct production of $D^{\pm}$ mesons from $c\bar{c}$ (and $B$ mesons from $b\bar{b}$) is charge symmetric. We also assume that there is negligible CP violation in the decays containing $D^{\pm}$ of $B^{\pm}$ mesons, or neutral $B^0_{(s)}$ mesons that have not oscillated into their antiparticle. We allow possible CP violation for $D^{\pm}$ mesons arising from the decay of oscillated $B^0_{d(s)}$ mesons, quantified by the mixing asymmetry parameters $a^{d(s)}_{\text{sl}}$ which are taken to be the current world averages $a^d_{\text{sl}} = (-0.09 \pm 0.21)\%$ and $a^s_{\text{sl}} = (-0.77 \pm 0.42)\%$~\cite{pdg}.

To determine the fraction of $D$ mesons in our sample that originate from such decays, we use MC simulation, passed through the full data reconstruction and reweighted to match the data in five important variables: the muon multiplicity, $p_T(\mu)$, $|\eta(\mu)|$, $q(\mu) \cdot q(D)$, and the separation of the muon and $D$ meson along the beam direction (at their point of closest approach in the transverse plane). The simulation is of $D^{\pm} \to K^{\mp} \pi^{\pm}\pi^{\pm}$ decays with the muon requirement only placed during simulation of the trigger and offline event selection, to ensure a representative mixture of muons from the initial hard scatter, from decays of heavy-flavor hadrons, and from decays of charged kaons and pions. A fraction $(52.3 \pm 0.3)$\% of $D$ mesons is found to originate from the decays of $B^0$ mesons, but only $(12.1 \pm 0.2)$\% from $B^0$ mesons that oscillated into their antiparticle prior to decay. For $B^0_s$ mesons, the corresponding fractions are $(2.7 \pm 0.1)$\% and $(1.33 \pm 0.06)$\%. Multiplying by the respective mixing asymmetries, the contributions to $A_{\text{phys}}$ are $(-0.010 \pm 0.023)$\% from $B^0$ and $(-0.004 \pm 0.002)$\% from $B^0_s$ mesons. The uncertainties are dominated by the $a^{d(s)}_{\text{sl}}$ inputs, and are taken as systematic. All other reasonable variations to the method (modified reweighting, adjusted lifetimes, mixing frequencies, and branching fractions) give negligible shifts with respect to the precision. Adding these contributions, we obtain $A_{\text{phys}} = (-0.014 \pm 0.023)$\%. Of the remaining $D$ mesons, $(35.9 \pm 0.3)$\% arise from direct $c\bar{c}$ hadronization, $(9.0 \pm 0.2)$\% are from $B^{\pm}$ decay, and the remaining $(0.10 \pm 0.02)$\% are from $b$ baryons. For all cases, the uncertainties on the quoted fractions come from the limited statistics of the simulation. 

From Eq.~(\ref{eq:main}), we obtain the final measurement
\begin{eqnarray}
A_{\text{\text{CP}}}(D^+ &\to& K^-\pi^+\pi^+)  \\
                  &=& [ -0.16 \pm 0.15 \text{\,(stat.)} \pm 0.09 \text{\,(syst.)} ]\%.\nonumber
\end{eqnarray}
In this evaluation, only the statistical uncertainty on $A$ is included in the final statistical uncertainty on $A_{\text{\text{CP}}}$. All other uncertainties are taken to be systematic, since they are not directly related to the size of the signal sample. They are added in quadrature and treated as completely uncorrelated, with the detailed breakdown given in Table~\ref{tab:syst}. This result is consistent with the standard model prediction of CP conservation.

\renewcommand\arraystretch{1.2}
\begin{table}[t]
\caption[]
{\label{tab:syst}Breakdown of the different sources of systematic uncertainty on the final $A_{\text{\text{CP}}}$ measurement, including contributions from the raw asymmetry $A$, from the kaon, muon, and pion inputs to the detector correction $A_{\text{det}}$, and from the physics asymmetry $A_{\text{phys}}$. All individual components are added in quadrature assuming zero correlations to obtain the total systematic uncertainty.}
\begin{center}
\begin{tabular}{|l|c|}
\hline \hline
Source                &  $\sigma_{\text{syst.}}(A_{\text{\text{CP}}})$    (\%) \\
\hline
Fit range ($A$)                   &  $0.017$ \\
Fit model ($A$)                   &  $0.005$ \\
Bin width ($A$)                   &  $0.005$ \\
Polarity weighting method ($A$)   &  $0.001$ \\
\hline
Kaon asymmetry statistics ($A_{\text{det}}$)   &  $0.040$ \\
Kaon asymmetry method ($A_{\text{det}}$)       &  $0.053$ \\
Muon asymmetry statistics ($A_{\text{det}}$)   &  $0.011$ \\
Muon asymmetry method ($A_{\text{det}}$)       &  $0.004$ \\
Pion asymmetry ($A_{\text{det}}$)              &  $0.050$ \\ 
\hline
$a^q_{\text{sl}}$ (dominates $A_{\text{phys}}$) & $0.023$  \\
\hline
Total                                 & $0.089$  \\
\hline \hline
\end{tabular}
\end{center}
\end{table}
\renewcommand\arraystretch{1.0}

We perform a range of cross-checks to demonstrate the stability of the measurement by repeating the entire analysis for orthogonal sub-samples of the data, divided in important variables including the LLR discriminant output, positive and negative kaon pseudorapidity, $p(K)$, $|\eta(K)|$, $q(\mu) \cdot q(D)$, and the instantaneous luminosity. In total, 19 such samples are tested, and all $A_{\text{\text{CP}}}$ measurements are consistent with the nominal value, with a $\chi^2$ of 13.6 for 12 degrees-of-freedom.

In conclusion, we have measured the direct CP-violating parameter in the Cabibbo-favored decay $D^+ \to K^- \pi^+ \pi^+$, finding an asymmetry consistent with the SM prediction of zero. The precision exceeds that of the previous best measurement by a factor of 2.5 and represents an important reference measurement for future studies of CP violation in charm and bottom hadron decays. In particular, it gives experimental confirmation of the assumptions used in measurements of CP violation in $D^0$ and $B^0$ mixing and decay~\cite{adsl,lhcb_deltaacp}, which is of special importance given the anomalously large asymmetry reported in same-charge dimuons~\cite{dimuon2014}, and for future searches for CP violation in bottom and charm hadrons.

%
We thank the staffs at Fermilab and collaborating institutions,
and acknowledge support from the
DOE and NSF (USA);
CEA and CNRS/IN2P3 (France);
MON, NRC KI and RFBR (Russia);
CNPq, FAPERJ, FAPESP and FUNDUNESP (Brazil);
DAE and DST (India);
Colciencias (Colombia);
CONACyT (Mexico);
NRF (Korea);
FOM (The Netherlands);
STFC and the Royal Society (United Kingdom);
MSMT and GACR (Czech Republic);
BMBF and DFG (Germany);
SFI (Ireland);
The Swedish Research Council (Sweden);
and
CAS and CNSF (China).
%

\end{document}